\documentclass[12pt,preprint]{aastex}

\newcommand{\be}{\begin{equation}}
\newcommand{\ee}{\end{equation}}

\def\lta{\,\raise 0.3 ex\hbox{$ < $}\kern -0.75 em
 \lower 0.7 ex\hbox{$\sim$}\,}
\def\gta{\,\raise 0.3 ex\hbox{$ > $}\kern -0.75 em
 \lower 0.7 ex\hbox{$\sim$}\,} 
\newcommand{\taueff}{\tau_{\rm eff}} 
\newcommand{\betacool}{\beta_{\rm cool}}  
\newcommand{\tcool}{t_{\rm cool}}  
\newcommand{\mjup}{M_{\scriptstyle \rm Jup}}
\newcommand{\mplan}{M_{\scriptstyle \rm P}}
\newcommand{\rdisk}{R_{\rm disk}}
\newcommand{\mdisk}{M_{\rm disk}}
\newcommand{\enhance}{\Lambda_{\rm ac}}
\newcommand{\enhancemax}{\Lambda_{\rm max}} 

\begin{document} 

\title{\bf Theoretical Mass Function for Secondaries Forming via \\
Gravitational Instability in Circumstellar Disks}

\author{Fred C. Adams,$^{a,b}$ Aster G. Taylor,$^{b,}$\footnote{Fannie
  and John Hertz Foundation Fellow} $\,$ and Michael R. Meyer$^b$}

\affil{$^a$Physics Department, University of Michigan, Ann Arbor, MI
  48109, USA}

\affil{$^b$Astronomy Department, University of Michigan, Ann Arbor, MI
  48109, USA}

\begin{abstract}
This paper constructs a theoretical framework for calculating the
distribution of masses for secondary bodies forming via gravitational
instability in the outer regions of circumstellar disks. We show that
several alternate ways to specify the mass scale of forming objects 
converge to the same result under the constraint that the parental
disks are marginally stable with stability parameter $Q=1$. Next we
show that the well-known constraint that the formation of secondary
bodies requires rapid cooling is equivalent to that of opacity limited
fragmentation. These results are then used to derive a mass function
for secondary objects forming through disk instablity. The resulting 
distribution is relatively narrow, with log-normal-like shape, a
characteristic mass scale of order $\mplan\sim10\mjup$, and an
approximate range of $4-80\mjup$. Current estimates for the occurrence
rate suggest that these objects are outnumbered by both stars and
planets formed via core accretion.
\end{abstract}


\section{Introduction}
\label{sec:intro}

Although {thousands of exoplanets and brown dwarfs} have been
detected, an empirical determination of their mass function has been
elusive, especially at lower masses, and a theoretical understanding
of the mass function remains in its infancy.  Compared to their
stellar counterparts, planets and brown dwarfs display a much wider
range of properties and are most likely formed through a variety of
mechanisms (for a recent review, see \citealt{drazkowska2023}). The
planetary population includes rocky worlds with relatively little
atmosphere and giant planets with masses dominated by their gas
content. Rocky planets are thought to form through the accumulation of
planetesimals and pebbles, whereas the gas giants primarily form
through a process of core accretion. At intermediate masses, the
planet population includes so-called superearths and sub-neptunes,
with varying abundances of rock, ices, and gaseous atmospheres. Due to
their formation mechanisms, one expects all of the aforementioned
planets to be enriched in heavy elements relative to their host stars.
In addition, however, some low-mass sub-stellar objects {(and
even stars)} can form through the action of gravitational
instabilities in circumstellar disks, and hence form with stellar
metallicity. {The goal of this present work is to calculate the
  distribution of masses for this collection of companions forming
  through gravitational instability in circumstellar disks.} The focus
is on Sun-like host stars, but the approach can be readily generalized
for other stellar masses. Although a great deal of previous work has
been done concerning the formation of secondaries via gravitational
instability, the expected distribution of masses (the secondary mass
function or SMF) has not been definitively determined. Toward that
end, {using a semi-analytic approach}, this paper constructs a
working theory for the SMF for this class of objects.\footnote{To
reiterate, this treatment considers secondaries that are formed
through gravitational instability {in disks to make up a
  particular} class of objects, with a well-defined mass
function. Rocky planets, giant planets forming through the core
accretion paradigm, and binary stellar companions thus represent
separate classes, each with their own mass distribution.}

In determining the secondary mass function, one can conceptually
divide the process into the following components. [1] First, one must
identify the mechanism(s) that determine(s) the mass $M$ of a forming
object. Specifically, we require a function of the form $M(x_1,x_2,\dots)$,
where the $x_k$ are the constituent variables. These quantities
specify the physical conditions that lead to gravitational collapse
and affect the mass of the secondary produced (e.g., disk surface
density $\Sigma$, sound speed $v_s$ which depends on disk temperature,
etc), but can also include stochastic variables.  [2] Next we must
determine or estimate the allowed distributions of the constituent
variables (e.g., the range of disk surface density over which
gravitational instabilities occur). [3] Finally, the distributions of
the constituent variables must be combined to construct the composite
distribution that is the SMF.

A large amount of previous work has been carried out regarding the
formation of giant planets and other secondaries by gravitational
instability (see the reviews of \citealt{durisen2007,kratter2016}).
Most of this work is numerical and consistently finds that the
secondary bodies have masses of order $M\sim10\mjup$ and form at large
radial distances from the star (e.g.,
\citealt{adams1992,boss1997,mayor2004,kratter2010,stamatellos2008};
\citealt{inutsuka2010,zhu2012}; and many others).  One recent set of
high-resolution simulations \citep{xu2024} finds a characteristic mass
scale of $M\sim20\mjup$, and a log-gaussian-like distribution over the
range $M\sim4-100\mjup$. A competing recent study \citep{boss2024}
produces a number of {secondary objects with smaller} masses
in the range $M=1-6\mjup$, with orbital radii in the range $r=30-50$
AU.  {Except for perhaps the latter study,} these results are
roughly consistent with previous findings that produce masses
$M=6-20\mjup$ and orbital distances $r\sim100$ AU \citep{boley2009},
and with population synthesis models
\citep{forgan2013,forgan2017,schib2023}, which produce objects with
mass $M=10-100\mjup$ and initial locations $r\sim100$ AU. In general,
numerical simulations show that gravitational instabilty leads to
secondary formation only in the outer disk, near $r\sim100$ AU
\citep{boley2009,vorobyov2013}. This circumstance arises due to the
requirement that gas must cool rapidly to form secondaries, and
cooling processes operate more effectively in the outer parts of the
disk (e.g., \citealt{rafikov2005,meru2010,lizano2010,forgan2011};
\citealt{kratter2011,boss2017}). The results must ultimately depend on
the parental circumstellar disks, which are likely to have a wide
range of poperties (e.g., \citealt{bate2018,andrews2020,tobin2020}).
However, unstable disks around Sun-like stars typically have masses
of order $M_d/M_\ast\approx0.1$ \citep{mann2015} and typical radii of
order $\sim100$ AU \citep{eisner2008,andrews2018}. Although the above
overview is incomplete, numerical simulations generally form
companions with masses of order $M\sim10\mjup$, and population
synthesis models have begun constructing the mass function
\citep{forgan2017,wagner2019,schib2023}. Nonetheless, a definitive
determination of the distribution of masses remains lacking, and this
paper seeks to gain additional insight into the problem with an
analytic approach. This work thus complements previous numerical
studies.

Although observations do not definitively distinguish between objects
formed by gravitational instability and those formed by other
mechanisms, a growing number of candidates have been detected.  With
expected masses $M\sim10\mjup$ and locations $r\sim100$ AU, these are
objects are best detected through direct imaging {and astrometry}. 
One recent survey using VLT data \citep{vigan2017} indicates that the
occurence rate of these secondaries (those with mass $M=1-75\mjup$ and
wide orbits of 20 -- 100 AU) falls in the range 1 -- 8\%. These
results are roughly consistent with more recent direct imaging surveys
(e.g., \citealt{nielsen2019,vigan2021}; and others), although the
total number of detections remains small. Note that secondary bodies
in the mass range of interest can overlap with planets formed by core
accretion on the low-mass end (say $M\lta15\mjup$) and brown dwarfs
{formed by star formation processes} on the high-mass end
($M\gta15\mjup$). {However, orbital separations provide some
discrimination: The occurence rate of Jovian planets peaks at
$a\sim3-5$ AU \citep{fulton2021,zink2023}, well inside the locations
of $\sim100$ AU expected for secondaries produced by disk instability.
For comparison, stellar binaries have a much wider range of
separations, from contact binaries to $a\gg100$ AU.}

This paper is organized as follows. The expected masses for objects 
forming through gravitational instability are derived in Section
\ref{sec:mass}. Here we find that a collection of different approaches
leads to the same effective mass scale, including the same functional
dependence on the background disk properties. The necessity of
sufficient cooling, taken up in Section \ref{sec:cooling}, constrains
the process of gravitational instability and provides an estimate for
the minimum secondary mass. We show that the resulting minimum mass is
equivalent to that obtained from the concept of opacity limited
fragmentation. An estimate for the maximum possible mass is given in
Appendix \ref{sec:maxfrag}. Using the mass scales and cooling
constraints, we construct a working theory of the secondary mass
function in Section \ref{sec:distrib}.  The paper concludes, in
Section \ref{sec:conclude}, with a summary of our results and a
discussion of their implications.

\section{Mass Scales for Secondary Formation} 
\label{sec:mass}

This section derives mass estimates for secondary bodies forming
through disk instability. The secondary mass, denoted here as $M$,
must arise from the circumstellar disk, which has surface density
$\Sigma$. If the formation occurs through gravitational instability,
then a region of area $A$ in the disk collapses to form the secondary.
The mass of the object, essentially by definition, must be given by 
\be
M = A \Sigma \,.
\label{basic} 
\ee
Note that the surface density $\Sigma$ is the value for the local
disk at the moment when the region collapses. This surface density,
in general, will be larger than the unperturbed value. We expect the
circumstellar disk to develop spiral structure so that the local
surface density within the spiral arms is larger than that of the
azimuthal average, and these regions of enhanced density are the
locations where objects could form. The enhancement factor is expected
to be greater than but of order unity, so that $\Sigma=\Lambda\Sigma_0$,
where $\Sigma_0$ is the unperturbed surface density. 

Equation (\ref{basic}) basically redefines the problem of determining
the secondary mass to that of determining the relevant area $A$. To
determine the area, we start by considering the relevant length scales
in the problem. These scales include the Hill radius $R_H$, the Bondi
radius $R_B$ \citep{bondi1952}, the disk scale height $H$, and the
length scale $\lambda_S$ for the most unstable spiral mode (e.g., 
\citealt{shubook}), and the Jeans length $\lambda_J$ \citep{jeans1902},
where these scales can be written 
\be
R_H = r \left({M \over 3M_\ast}\right)^{1/3} \,,
\qquad
R_B = {GM \over v_s^2} \,,
\qquad 
H = {v_s \over \Omega} \,,
\qquad 
\lambda_S = {v_s^2 \over G\Sigma}\,,
\qquad
\lambda_J = {\sqrt{\pi}v_s\over\sqrt{G\rho}} \,.
\label{scales} 
\ee
In addition to the mass $M$ of the secondary, these scales depend on
the radial location $r$ of the object, the local sound speed $v_s$,
the mean motion $\Omega$ of the orbit, the stellar mass $M_\ast$, the
local disk surface density $\Sigma$, the gravitational constant $G$,
and the local density $\rho$.  This section determines the mass scales
for forming secondary bodies where the area in equation (\ref{basic})
is determined by each of these length scales. For all of the cases,
the disk must be gravitationally unstable, which in turn requires that
the Toomre stability parameter $Q$ must be of order unity
\citep{toomre1964}, where
\be
Q = {v_s \kappa \over \pi G \Sigma} \rightarrow
{v_s \Omega \over \pi G \Sigma} \,. 
\label{qtoomre} 
\ee
Throughout this work, we assume that the rotation curve $\Omega$ is
close enough to Keplerian that we can use $\Omega$ in place of the
epicyclic frequency $\kappa$.

\subsection{Mass Defined by the Hill Radius}
\label{sec:hill} 

To find the mass scale defined by the Hill radius $R_H$,
we let the area $A$ be given by 
\be
A = \xi \pi R_H^2 \,,
\ee
where $\xi$ is a dimensionless parameter of order unity.  Its value
can be varied to account for different (non-circular) geometry. In
addition, the value of $\xi$ could have a stochastic character. The
mass is given by $M=A\Sigma$, but the area depends on the Hill radius,
which in turn depends on the mass. We must solve for the secondary
mass to find 
\be
M = {(\xi \pi r^2 \Sigma)^3 \over 9 M_\ast^2} \,. 
\ee
We can get rid of $\Sigma$ in favor of $Q$ using the
stability criterion of equation (\ref{qtoomre}) to write 
\be
M_{\scriptstyle\rm Hill} = {1\over9}
\left({\xi\over Q}\right)^3 {v_s^3 \over G\Omega} \,.
\label{masshill} 
\ee

\subsection{Mass Defined by the Bondi Radius}
\label{sec:bondi} 

If we use the Bondi radius to define the area $A$, with all of
the other parameters the same, the mass scale takes the form 
\be
M = \xi\pi \left({GM\over v_s^2}\right)^2 \Sigma\,,
\ee
which can be solved to find
\be
M = {v_s^4 \over \xi\pi G^2 \Sigma} \,.
\ee
Using the expression for $Q$ we find
\be
M_{\scriptstyle\rm Bondi} = {Q\over\xi} {v_s^3 \over G\Omega} \,.
\label{massbondi} 
\ee

\subsection{Mass Defined by the Scale Height}
\label{sec:height} 

Similarly, for the scale height we find the mass scale 
\be
M = \xi\pi \left({v_s\over\Omega}\right)^2 \Sigma\,,
\ee
which becomes
\be
M_{\scriptstyle\rm Height} 
= {\xi\over Q} {v_s^3 \over G\Omega} \,.
\label{massheight} 
\ee

\subsection{Mass Defined by the Spiral Structure Length Scale}
\label{sec:spiral} 

For the length scale $\lambda_S$, defined through the dispersion
relation for spiral density wave theory (e.g., \citealt{shubook}),
we find 
\be
M = \xi \pi \lambda_S^2 \Sigma = \xi \pi 
{v_s^4 \over G^2 \Sigma}\,. 
\ee
Using the Toomre $Q$ parameter to remove the dependence on
surface density, the mass scale becomes 
\be
M_{\scriptstyle\rm Spiral} = \xi \pi^2 Q {v_s^3 \over G\Omega} \,.
\label{massspiral} 
\ee
Note that for $Q=1$, the length scale $\lambda_S=2\pi{H}$. As a
result, the convergence of the mass scales given in equations 
(\ref{massheight}) and (\ref{massspiral}) is expected, up to factors
of order unity (see also \citealt{boley2010,forgan2011,kratter2016}). 

\subsection{Mass Defined by the Jeans Length}
\label{sec:jeans}

Next we use the Jeans length \citep{jeans1902} to define the relevant
area, so that the mass scale takes the form
\be
M_{\scriptstyle\rm Jeans} = \xi \lambda_J^2 \Sigma \,,
\ee
where $\lambda_J$ is the Jeans length. 
The density $\rho$ is given by 
\be
\rho = {\Sigma \over 2H} 
\qquad \Rightarrow \qquad
\lambda_J = v_s^{3/2}
\left({2\pi\over G\Sigma\Omega}\right)^{1/2}\,. 
\label{rhomega} 
\ee
The mass scale from Jeans arguments thus becomes  
\be
M_{\scriptstyle\rm Jeans} = 2\pi\xi {v_s^3 \over G\Omega} \,.
\label{massjeans} 
\ee

\subsection{Synthesis and Dimensional Analysis}
\label{sec:dimension}

The above arguments considered five seemingly different ways to define
a mass scale for secondary bodies collapsing out of an unstable
circumstellar disk. However, the above derivations show that all five
results can be written in the same simple form 
\be
M = \eta {v_s^3 \over G \Omega} \,, 
\label{massgeneral} 
\ee
where the dimensionless coefficient $\eta$ is of order unity. This
coefficient depends on the geometrical factor $\xi$ and the stability
parameter $Q$, which are both dimensionless and of order unity. This
section shows that this convergence is expected.

The scales in the problem include the mass $M$, the local mean motion
$\Omega$, the location $r$ of the forming object, the gravitational
constant $G$, the sound speed $v_s$, the disk surface density
$\Sigma$, the disk scale height $H$, the disk volume density $\rho$,
and the stellar mass $M_\ast$. In addition, all of the dimensionless
factors in the problem can be encapsulated by the variable $\eta$,
which will have some (as yet unspecified) distribution of values (but
will be of order unity).  We thus want to determine the function
$M(r,\Omega,G,v_s,\Sigma,H,\rho,M_\ast,\eta)$, where all of the
variables could in principle play a role in secondary formation.

The variable list can be reduced as follows. Since we include the mean
motion $\Omega$, surface density $\Sigma$, and density $\rho$, the
orbital location $r$ only enters through tidal effects, which
determine the Hill radius $R_H=r(M/3M_\ast)^{1/3}$.  However, this
expression can be rewritten as $R_H = (GM/3\Omega^2)^{1/3}$. As a
result, the stellar mass $M_\ast$ and orbital location $r$ can be
eliminated from the variable list (since the list already includes
$G$, $M$, and $\Omega$). This argument assumes that non-local (i.e.,
stellar) environmental effects only enter the problem as tidal
forces.\footnote{This result can be understood as follows: Since the
forming object is small (in mass) compared to the star, one can
consider the parcels of gas that eventually collapse to form the
object as test particles in the restricted three-body problem.
Specifically, since $M\ll M_\ast$, we can work in the Hill regime, so
that the additional `forces' acting on the collapsing parcels of gas
are the Coriolis forces ($\sim\Omega v$) and effective tidal forces
($\sim\Omega^2r_\perp$). In the reference frame of the forming object,
all of the environmental effects are thus encapsulated via the
parameter $\Omega$.}

Next we note that the disk scale height $H=v_s/\Omega$ and the volume
density $\rho$ = $\Sigma/2H$ = $\Sigma\Omega/2v_s$. As a result, we
can eliminate $H$ and $\rho$, and are thus left with only the
variables $(M,\Omega,G,v_s,\Sigma)$. Finally, the requirement that the
disk must be unstable implies that the Toomre $Q$ parameter must be of
order unity, which in turn allows us to eliminate another 
variable.\footnote{The stability parameter must be near $Q\approx1$.
If $Q$ is larger, then gravitational instability does not occur. But
$Q$ can never be significantly smaller than unity. The disks are
generally stable to begin with, so that they must pass through
$Q\sim1$ in order to reach $Q<1$, which means that they will fragment
near $Q=1$.} Here we eliminate the surface density $\Sigma$ in favor
of $Q$, and are left with only four independent variables
$(M,\Omega,G,v_s)$.  Dimensional analysis shows that one can construct
one and only one dimensionless field (denoted here as $\eta$) from
these variables, so we can write
\be
\eta \equiv {GM\Omega \over v_s^3} \qquad {\rm or} \qquad
M = \eta {v_s^3 \over G\Omega} \,.
\ee
With only one dimensionless field, $\eta$ must be constant and is
expected to be of order unity. We thus obtain the same expression for
the mass scale found previously.

As shown below, the following picture emerges: Secondaries forming via
gravitational instability must collapse at orbital locations of order
$r$ = 100 AU.  If the surface density perturbations are moderately
nonlinear, the disk mass must be of order $\mdisk=0.1M_\odot$. With a
typical temperature of $T=35$ K, the quantity 
$v_s^3/G\approx0.01\mjup$ yr$^{-1}$. Since the orbit time at 100 AU is
of order $P$ = 1000/($2\pi$) yr, and the dimensionless parameter
$\eta\sim5-10\sim2\pi$, the resulting mass scale is $M\sim10\mjup$.

\section{Cooling Constraints}
\label{sec:cooling} 

The mass scales of the previous section correspond to the mass that
an object would have if it forms via gravitational collapse within a
disk, provided that it can actually form. In order to produce a
secondary --- and have the object survive --- the collapsing gas must
be able to cool sufficiently rapidly.  This constraint is usually
written in terms of a cooling time criterion of the form 
\be
\tcool \Omega < \betacool \,,
\label{cooling} 
\ee
where the dimensionless parameter $\betacool\approx3$ (e.g.,
\citealt{gammie2001}). The cooling time can be written in the form
\be
\tcool = {v_s^2 \Sigma \over \sigma T^4} \taueff\,,
\label{tcool} 
\ee
where the effective optical depth is defined as 
\be
\taueff \equiv {\tau + 1/\tau \over 2} 
\qquad {\rm where} \qquad \tau = \kappa_R \Sigma \,,
\ee
and where $\kappa_R$ is the Rosseland mean opacity. This section
reviews how the cooling constraint of equation (\ref{cooling})
places corresponding constraints on the disk surface density, 
disk mass, formation radius, and fragmentation mass.

\subsection{Surface Density Constraint}
\label{sec:surfconstraint}

The cooling constraint implies a corresponding constraint on the disk
surface density required for secondary formation.  Specifically, the
requirements $\Omega\tcool<\betacool$ and $Q<1$ jointly imply the
coupled inequalities 
\be
{\Omega \Sigma \taueff \over \betacool \sigma}
\left({k\over\mu}\right)^4 < v_s^6 < 
\left({\pi G \Sigma \over \Omega}\right)^6\,. 
\ee
Eliminating the central expression, one obtains a constraint on
the surface density, which can be written 
\be
\Sigma > \left(
{\Omega^7 \taueff\over 2\pi^6 \betacool G^6 \sigma}
\right)^{1/5} \left({k\over\mu}\right)^{4/5} \,. 
\ee
If we set $\taueff=1$ and $\betacool=3$, and take $M_\ast$ = 1
$M_\odot$, this surface density constraint evaluates to become 
\be
\Sigma > 20 \, {\rm g} \, {\rm cm}^{-2}
\left( {r \over 100\,\,{\rm AU}}\right)^{-21/10} \equiv
\Sigma_{\rm min} \,.
\label{surfcon} 
\ee
Keep in mind that this surface density is that of the local area that
collapses to form a secondary. The azimuthally averaged surface density
$\Sigma_0$ could be lower by a factor $\Lambda$ (so that $\Sigma$ =
$\Lambda\Sigma_0$). 

\subsection{Disk Mass Constraint}
\label{sec:diskmass}

The surface density constraint implies a corresponding limit on the
disk mass. If we define $\xi=r/\rdisk$ and let $\Sigma_{100}$ be the
surface density at $r=\rdisk=100$ AU, the minimum disk mass is given by 
\be
\mdisk = 2\pi \rdisk ^2 \Lambda^{-1} \Sigma_{\rm disk}
\int_0^1 \xi d\xi \xi^{-p} \,,
\ee
where the power-law index $p$ is expected to lie in the range $1<p<2$,
with some preference for $p\approx3/2$ at the early stages when we
expect gravitational fragmentation to occur. For the choice $p=3/2$
and $\rdisk=100$ AU, the minimum disk mass is $\mdisk\approx$ 0.28
$\Lambda^{-1}$ $M_\odot$. Note that a disk mass $\mdisk\sim0.3M_\odot$
is somewhat larger than those observed and larger than that expected
due to considerations of gravitational instability \citep{shu1990},
where we expect $\mdisk/M_\ast\sim0.1$. A modest amplification factor
$\Lambda\gta3$, however, implies that the disk mass can exceed the
minimum and still be consistent with expectations. Notice also that
if the index $p=1$, the required disk mass decreases to only
$\mdisk\approx$ 0.14 $\Lambda^{-1}$ $M_\odot$.

\subsection{Radial Location Constraint}
\label{sec:location}

For completeness we note that the cooling constraint implies that
gravitational instability must preferentially produce objects in the
outer disk. Combining equations (\ref{cooling}) and (\ref{tcool}),
we find 
\be
\betacool > {v_s^2 \Sigma \Omega \over \sigma T^4} \taueff
> {v_s^2 \Sigma_{\rm min} \Omega \over \sigma T^4} \propto
r^{-p - 3/2 + 3q}\,. 
\ee
The final expression gives the radial dependence of the expression,
where $(p,q)$ are the indices of the surface density and temperature
distributions, and where the rotation curve $\Omega$ is Keplerian
(since $M_d\ll M_\ast$). For $\betacool=3$ and the minimum surface
density of equation (\ref{surfcon}), this constraint evaluates to 
\be
r \gta 20 \, {\rm AU}\,
\left({T_{100} \over 30\,{\rm K}}\right)^{-10/7} \,,
\ee
where $T_{100}$ is the disk temperature at 100 AU. Keep in mind that
this constraint represents a lower limit to the formation location,
i.e., the typical formation location will be farther from the star.
This constraint indicates that secondary formation must take place at
distances roughly comparable to the expected disk sizes. As a result,
such objects are expected to form over a relatively narrow range of
radii of order 100 AU.

Note that the combination of a minimum secondary mass, a minimum disk
surface density, and a minimum radial location is highly constraining.
Circumstellar disks tend to be more massive during the early formative
epochs, but then their radial sizes are somewhat smaller. Later on,
disks found in association with T Tauri stars are often $\sim100$ AU
in radius, but their masses are rarely large enough for gravitational
instability to occur (see the review of \citealt{andrews2020} for a
summary of the observations). These combined constraints thus imply
low occurence rates for secondaries formed via gravitational
instability.

\subsection{Minimum Fragmentation Mass}
\label{sec:minfrag} 

These cooling considerations imply a minimum mass for bodies that form
through gravitational fragmentation of disks. This argument is
essentially the same as the traditional derivation of the minimum mass
given by opacity limited fragmentation \citep{rees1976}, where this
argument has been revived and extended many times (e.g.,
\citealt{rafikov2005,kratter2010,lizano2010,forgan2011}).

If we use the cooling constraint of equation (\ref{cooling})
and the constraint that $Q<1$, we have
\be
\betacool \sigma T^4 > v_s^2 \Omega\Sigma \taueff >
v_s^2 \Omega\Sigma > {v_s^3 \Omega^2 \over \pi G} \,. 
\ee
For convenience (following \citealt{rees1976}), we define an
effective gravitational fine structure constant 
\be
\alpha_G \equiv {G \mu^2\over \hbar c} \approx 3.1 \times
10^{-38} \,, 
\ee
where $\mu$ is the mean mass of the particles in the gas
(expected to be $\mu\approx2.3m_P$ for solar composition).
With this definition, the above inequality becomes
\be
{v_s^6 \over G^2 \Omega^2} > {v_s\over c}
{60\over\pi^3\betacool} \alpha_G^{-3} \mu^2 \,.
\ee
Using our results for the mass of the fragments we find
\be
M_{\rm min} = \eta {v_s^3 \over G\Omega} > \eta
\left({v_s\over c}\right)^{1/2}
{60\over\pi^3\betacool} \alpha_G^{-3/2} \mu \approx
\eta \left({v_s\over c}\right)^{1/2}(224 \mjup)\,.
\ee
If we take $\eta=10$ and the minimum possible sound speed
($v_s=0.2$ km/s for $T$ = 10 K), we find
$M_{\rm min} \approx 2 \mjup$. The exact value depends on
the choice of the dimensionless constants. For example, Whitworth
and Stametellos (2006) find $M_{\rm min}\approx4\mjup$.
Note that the minimum mass scale is essentially just the
Chandrasekhar mass multiplied by the factor $(v_s/c)^{1/2}$.
{Note that this mass scale does not depend on the mass of
the host star.}  

Now let's compare the cooling argument in a disk to the original
calculation \citep{rees1976} in a molecular cloud. The cooling
argument in a disk implies 
\be
\sigma T^4 > {1\over3} \taueff \Omega v_s^2 \Sigma >
{1\over3} \Omega v_s^2 \Sigma \,,
\ee
whereas the requirement for a collapsing cloud to be able
to radiate its energy has the form
\be
\sigma T^4 > {GM^2 \over 4\pi r^3} \sqrt{G\rho} =
{1\over3} M (G\rho)^{3/2} \,. 
\ee
For unstable disks with $Q=1$, we know that $G\rho$ =
$\Omega^2/(2\pi)$ from equation (\ref{rhomega}).  If we ignore
dimensionless constants of order unity, the right hand side of the
Rees constraint becomes $M\Omega^3$.  If we use $Q=1$, and the mass
formula $M\sim v_s^3/(G\Omega)$, the right hand side of the disk
constraint also becomes $M\Omega^3$. The two constraints are thus the
same up to dimensionless factors of order unity. In other words, the
minimum mass for secondary formation in disks due to cooling and
gravitational instability is simply the previously-derived mass scale
for opacity limited fragmentation (in somewhat re-packaged form).

\section{The Secondary Mass Distribution}
\label{sec:distrib}

This section constructs the mass distribution for objects forming
through gravitational instabilities. We start with the expression
for the secondary mass in terms of constituent variables, 
\be
M = \eta {v_s^3 \over G\Omega} \,.
\label{massone} 
\ee
If the distributions of the individual variables $(\eta,v_s,\Omega)$
were known, and if they were independent, then we could directly
calculate the distribution of the composite variable, here the mass
$M$ (see \citealt{adams1996} and \citealt{adams2021} for analogous
treatments for the mass function of forming stars and giant planets
forming through core acccretion). The first step is thus to specify 
the individual distributions. 

\subsection{Scaling Considerations}
\label{sec:scaling}

The variables appearing in equation (\ref{massone}) are not
independent. Both the sound speed (via the temperature profile) and
the rotation curve are functions of radius, and are thus correlated.
In other words, we cannot sample the sound speed $v_s$ (which depends
on radial position) independently of the rotation rate $\Omega$ (which
also depends on position). In order to construct independent variables
to sample, we must consider how the constituent variables depend
on radius, as well as stellar mass.

Within a given circumstellar disk, the temperature profile is expected
to have a power-law form 
\be
T_d (r) = T_D \left({R_D \over r}\right)^q \,, 
\ee
where $q\approx1/2-3/4$ and $R_D$ is the outer disk radius
{\citep{shu1990,hartmann2009}}. For the usual choice $q=1/2$
(e.g., \citealt{andrews2007}), the cofficient $T_D$ is given by 
\be
\sigma T_D^4 = {1 \over \ln(R_D/R_\ast)}
{L_D \over 4\pi R_D^2} \,. 
\ee
{In addition to viscous accretion, } 
the total effective disk luminosity includes the power that is
intercepted and reprocessed from the star, and that contribution
represents a fraction ${\cal F}=1/4-1/2$ of the stellar luminosity.
For pre-main-sequence stars, the mass-luminosity relation takes
the approximate form $L_\ast\sim M_\ast^2$, where this expression
follows from fitting the results from stellar evolution simulations
\citep{paxton2011} for stars of varying mass and constant PMS ages.
The scaling law for the varaible $T_D$ then takes the form 
\be
T_D \propto M_\ast^{1/2} \,. 
\ee
The rotation curve of the disk is nearly Keplerian, so that
$\Omega \propto M_\ast^{1/2}$. These results produce a weak scaling
of the fragment mass scale with the stellar mass, where this law
can be written in the form  
\be
M \sim M_\ast^{1/4} \,. 
\ee
Observations indicate that smaller stars (red dwarfs) have lower
occurence rates for giant planets (although more data are necessary to
define the trends; see, e.g., \citealt{bryant2023}). In any case, the
stellar mass range of interest is limited, perhaps to a factor of
$\sim4$, which contributes a factor of $\sqrt{2}$ to the range of
secondary masses.  With the above specifications for the temperature
distribution, and hence the sound speed, the secondary mass $M$ scales
with radius according to
\be
M \sim {T_d^{3/2} \over \Omega} \sim r^{3/4} \,. 
\ee
The secondary mass {varies somewhat less than linearly with} the
location $r$. Note that the objects are expected to migrate both
inward and outward from their birth locations. As a result, the range
of possible radii contributes to the width of the mass function, and
is included here (see below), but we assume the objects move enough
that the resulting distribution is independent of the final semimajor 
axis. {In this treatment, we calculate a single mass distribution
for the entire population of secondaries (with expected separations
of $\sim30-300$ AU). Future generalizations of this work should
determine the SMF as a function of final location of the objects.}

The scaling laws outlined above account for the variations in the
radial location of the forming object and the mass of the host
star. Since both the sound speed $v_s$ and the mean motion $\Omega$
depend on both $r$ and $M_\ast$, we need to include the above scalings
in order to keep the sampling of variables independent. With the mass
and position determined, the rotation rate $\Omega$ is specified.
However, the temperature profile, equivalently the distribution of
sound speed, can include additional variations from source to
source. Even after accounting for the crude scaling of the coefficient
$T_D$ with stellar mass (over a relatively narrow mass range), we
expect additional system-to-system variations. With these
considerations in mind, we define the following suite of dimensionless
variables: 
\be
\eta = \eta \, \qquad \xi = {r \over 100 AU}\,, \qquad
\mu = {M_\ast\over1M_\odot} \,, \qquad {\rm and} \qquad
\beta = {T_D \over T_{D0}} \,,
\ee
where the benchmark value of disk temperature is independent of
stellar mass. In other words, we are assuming that the dependence of
the distribution of temperature coefficient $T_D$ has a mass dependent
part and an additional variation encapsulated through the variable
$\beta$. With these definitions, the secondary mass formula has the
form
\be
M = M_0 \eta \beta^{3/2} \mu^{1/4} \xi^{3/4} \,,
\label{scalemass} 
\ee
where the mass scale $M_0$ is calculated using the benchmark
values of the dependent variables.

Next we put this expression in a more mathematically
convenient form. First note that
\be
\ln(M/M_0) = \ln\eta + {3\over2} \ln\beta + {1\over4}
\ln\mu + {3\over4}\ln\xi \,.
\label{logform} 
\ee
Let us define new variables
\be
x_\eta = \ln\eta \,, \qquad
x_\beta = \ln\beta \,, \qquad
x_\mu = \ln\mu \,, \qquad {\rm and} \qquad 
x_\xi = \ln\xi \,.
\ee

Note that we can take the variables $x_k$ to have zero mean. To
illustrate this point, consider the first variable $\eta$.  The
variable $\eta$ will have a distribution, so the corresponding
logarithmic variable $\ln\eta$ will have a distribution, which
can be denoted as $f_\eta(x_\eta)$. The mean value is given by 
\be
\langle{x_\eta}\rangle = \langle{\ln\eta}\rangle =
\int_{-\infty}^\infty \ln\eta f_\eta(\ln\eta)d\ln\eta\,.
\ee
If the mean $\langle{\ln\eta}\rangle$ is non-zero, then we
can define a new variable $\eta'$ such that
\be
\ln\eta' \equiv \ln\eta - \langle{\ln\eta}\rangle
\qquad \Rightarrow \qquad \eta' =
\eta\,{\rm e}^{-\langle{\ln\eta}\rangle}\,.
\ee
The constant $\langle{\ln\eta}\rangle$ can be incorporated
into the definition of the benchmark mass scale $M_0$. 
We can thus use the expression of equation (\ref{logform})
with the variables having zero mean. 

We thus have a composite variable that is the sum of
four constituent variables:
\be
\ln(M/M_0) = \zeta \qquad {\rm where} \qquad
\zeta = x_\eta + {3\over2} x_\beta + {1\over4}
x_\mu + {3\over4}x_\xi \,.
\ee
Note that $\zeta$ has zero mean. Since the constituent variables (here
the $x_k$) are constructed so that they are statistically independent
(and since $\langle\zeta\rangle$ = 0), the variance of the composite
variable is given by
\be
\sigma_\zeta^2 =
\sigma_\eta^2 + {9\over4}\sigma_\beta^2
+ {1\over16} \sigma_\mu^2 + {9\over16} \sigma_\xi^2 \,.
\label{totalwidth} 
\ee
This expression thus defines the width of the secondary mass
function for secondaries forming via gravitational instability. 

\subsection{Width of the Composite Distribution}
\label{sec:width} 

If a variable $x$ varies by a factor of $\Lambda$ over its range, so
that it varies by a factor of $\sqrt{\Lambda}$ in either direction
from its geometric mean, and if the distribution is log-uniform, then
the variance is given by 
\be
\sigma^2 = {(\ln\Lambda)^2 \over 12} \,.
\label{sigmax}
\ee
To fix ideas, consider the allowed range of radial locations to vary
from $r=30$ AU to $r=300$ AU, so that $\Lambda_\xi=10$. {The lower
  limit arises from the radial location constraint of Section
  \ref{sec:location}, and the upper limit arises from observations
  (e.g., \citealt{eisner2008,andrews2020}).} The temperature cannot
be much colder than 35 K due to the background radiation from the
cluster, but at the large radii necessary for gravitational
instability (see the previous section), the temperature cannot be too
large {(e.g., see Figure 2 of \citealt{kratter2016})}, so we estimate
the range to be $\Lambda_\beta=\sqrt{10}$. The stellar mass could also
vary by a factor of $\Lambda_m=10$.  The range in $\eta$ is less well
determined, {but the resulting masses must be larger than the
  opacity limited fragmentation (Section \ref{sec:minfrag}), and
  smaller than the total disk mass (Appendix \ref{sec:maxfrag}). As a
  result, we take the range for $\eta$ to be} $\Lambda_\eta=10$. With
these estimates, the total width of the distribution is given by 
\be
\sigma_\zeta^2 = {(\ln10)^2 \over 12} \left[ 1 
+ {9\over16} + {1\over16} + {9\over16} \right] =
{35\over192}(\ln10)^2 \approx 0.97\,.
\label{zetasig}
\ee
The variance is thus close to unity. We can define an `effective mass
range' for the composite distribution by inverting equation
(\ref{sigmax}), i.e., find the value of $\Lambda$ that has the same
variance (\ref{zetasig}) as the composite distribution. The resulting
estimate for the range metric is given by 
\be
\Lambda_\zeta = \exp[ 2 \sqrt{3} \sigma_\zeta] \approx 30\,.
\label{zetalamb} 
\ee
If the mass coefficient $M_0\sim16\mjup$, for example, the resulting
range of secondary masses would vary from 3 to 90 $\mjup$, which
is reasonably close to the estimated range from the minimum (Section
\ref{sec:minfrag}) and maximum (Appendix \ref{sec:maxfrag}) mass
scales calculated independently. Note that equation (\ref{zetalamb})
is a measure of the effective range, rather than the full range over
which the distribution is nonzero.

\begin{figure}
\centering
\includegraphics[width=0.80\linewidth]{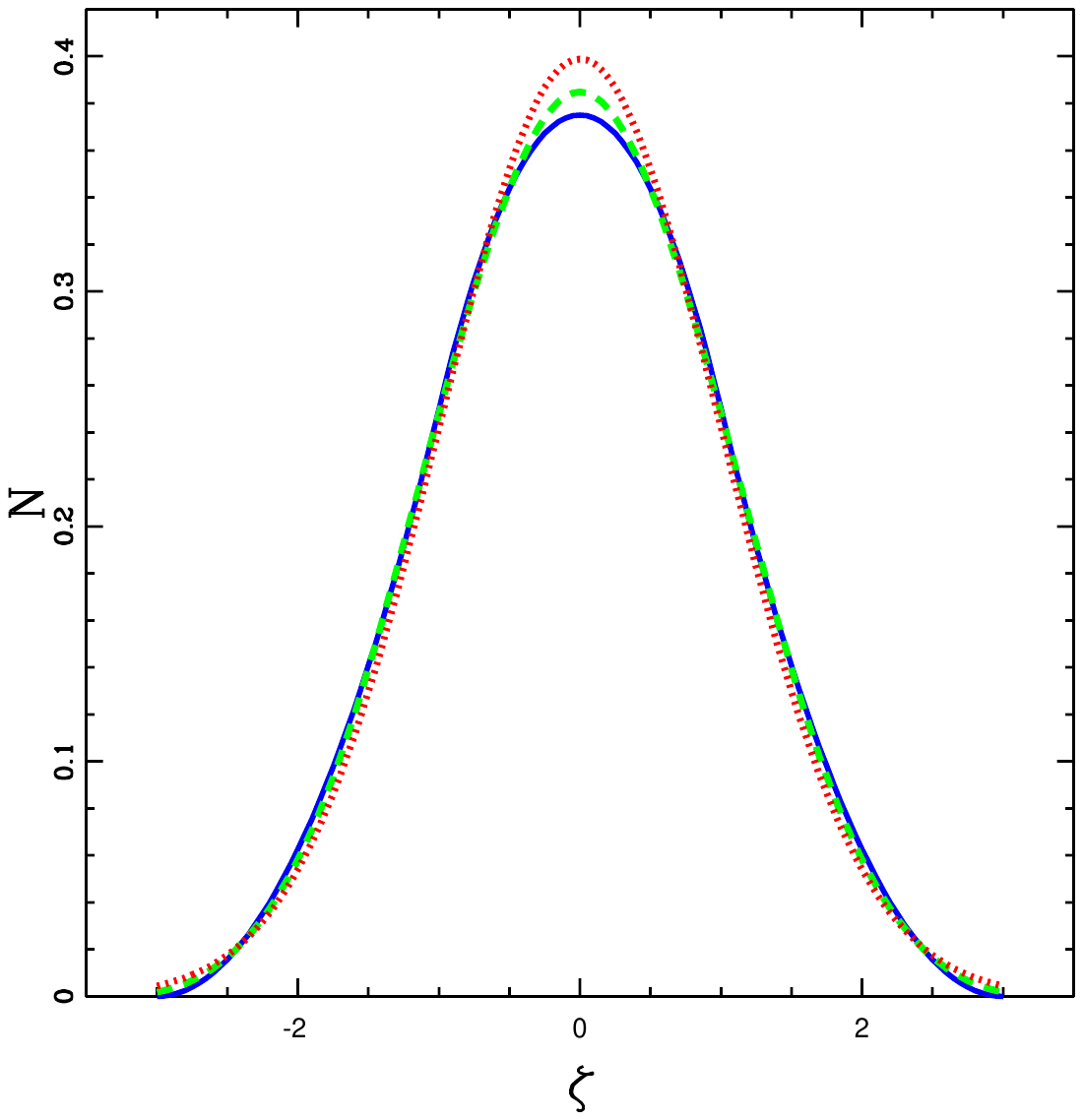}
\vskip-1.30truein
\caption{Comparison of secondary mass distributions in terms of
  $\zeta=\ln(M/M_0)$.  The three curves correspond to the sum of
  three random variables ($f_3$, solid blue), four random variables
  ($f_4$, dashed green), and the limiting case of an infinite number
  of random variables (a gaussian, dotted red). All of the
  distributions have the same total variance and zero mean. Given the
  small number of objects in the observed distribution, all of these
  functional forms are equivalent. Conceptually, however, the limiting
  case of the gaussian form has non-zero support at large values of
  $\zeta$, whereas the cases of $f_3$ and $f_4$ (and the physical
  distribution) have a strictly finite extent. }
\label{fig:dist}
\end{figure} 

\subsection{Secondary Mass Distribution}
\label{sec:massdist} 

The considerations of the previous section show that the overall width
of the mass distribution is given by $\sigma_\zeta\approx1$ (see
equation [\ref{zetasig}]). Although the composite variable $\zeta$ is
the sum of four constituent variables, the result is only weakly
dependent on the mass variable, whereas the other three contribute
more or less equally. As a result, we can model the composite variable
$\zeta$ (equivalently $\ln[M/M_0]$) as the sum of three identically
distributed random variables with individual distributions that are
uniformly distributed in logarithmic space, i.e, 
\be
p(x) = {1\over2w} \qquad {\rm for} \quad x \in [-w,w]\,,
\label{unilog} 
\ee
so that the individual variances are $\sigma_1^2 = w^2/3$.  Since
$\sigma_\zeta^2=\sigma_\zeta=1$, we need the widths of the individual
distributions to have the value $w=1$. In general, the composite
distribution for the sum of three (identical) random variables is
given by the expression 
\be
f_3 = {dP\over d\zeta} = {1\over16w^3}\left\{
\matrix{2(3w^2 - \zeta^2) & {\rm for} & |\zeta|<w \cr
  (|\zeta|-3w)^2 & {\rm for} & w < |\zeta|< 3w \cr
   0 & {\rm for} & |\zeta|> 3w} \right.\,.
\ee
This expression can be derived by convolving three copies of
the distribution of equation (\ref{unilog}). 
The mass function itself is then given by
\be
{dP\over dM} = {dP\over d\zeta} {d\zeta\over dM} =
{1\over M} {dP\over d\zeta} \,.
\ee
One can show that the peak of the mass function occurs
at the value
\be
\zeta_\ast = 1 - \sqrt{1 + 3w^2} \,, 
\ee
for arbitrary values of $w$. Here $w=1$, so the peak in
$dP/dM$ occurs at $\zeta_\ast=-1$, i.e., at mass $M=M_0$/e
$\approx$ 0.368 $M_0\sim6\mjup$ {(for our standard
values of parameters).} 

For completeness, note that for the case of the sum of four
random variables, the composite distribution takes the form
\be
f_4 = {dP\over d\zeta} = {1\over96w^4}\left\{
\matrix{32 w^3 - 12 w \zeta^2 + 3|\zeta|^3 & {\rm for} & |\zeta|<2w \cr
(4w-|\zeta|)^3 & {\rm for} & 2w < |\zeta|< 4w \cr
0 & {\rm for} & |\zeta|> 4w} \right.
\ee

We can compare the distributions with three and four random variables,
as shown in Figure \ref{fig:dist}. Also shown is the gaussian form
that results from taking the limit of an infinite sum of random
variables, i.e., from the Central Limit Theorem (e.g.,
\citealt{richtmyer1978}). All three composite distributions are chosen
to have the same variance and zero mean. Significantly, all three
distributions are nearly the same. The largest difference between the
finite variable cases and the gaussian limit occurs in the tails of
the distribution. For a finite number of variables, the distributions
vanish beyond a finite range of the composite variable (here,
$\zeta$), whereas the gaussian only approaches zero in the limit
$\zeta\to\infty$. Since we expect objects formed via gravitational
instability to have a finite range, the distributions with finite
width (here $f_3$ and/or $f_4$) provide better models for the SMF.
Strictly speaking, the gaussian limit is `infinitely wrong' for masses
outside the allowed range $M_{\rm min}\le{M}\le{M}_{\rm max}$.

{Given that a wide range of distributions are possible, one might
worry that the nearly-gaussian form of the composite distribution
depends sensitively on our choice of a log-uniform distribution for
the constituent variables (see equation [\ref{unilog}]). Appendix
\ref{sec:altprob} shows that different choices for the constituent
distributions produce similar results for the composite distribution,
and quantifies possible departures from a gaussian form (see Figure
\ref{fig:altprob}). The SMF is thus expected to be relatively close
to log-normal.}\footnote{In the limit where the number of
independent constituent variables is large, the Central Limit
Theorem shows that the composite distribution approaches a normal
form. Here we find that the result is relatively close to gaussian
even with only $N$ = 3 and 4 variables.} 

\begin{figure}
\centering
\includegraphics[width=0.80\linewidth]{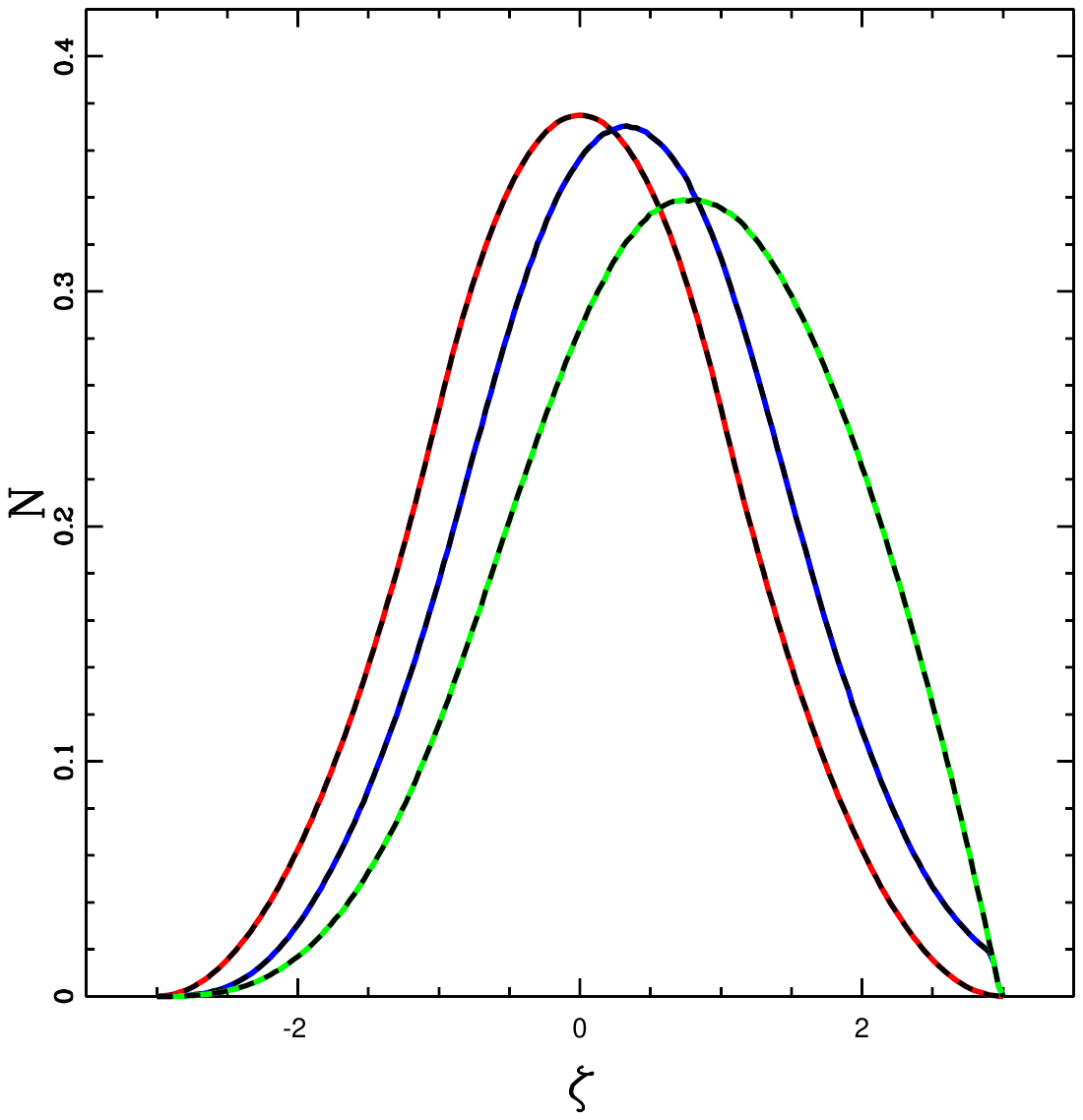}
\vskip-1.30truein
\caption{Effects of continued accretion on the secondary mass
  distribution. The left (red) curve shows the initial distribution
  (without accretion). The middle (blue) curve shows the mass
  distribution for the case with a log-uniform distribution of the
  enhancement factor with a maximum of $\enhancemax=2$. {The right
  (green) curve shows the mass distribution with continued mass
  accretion with a maximum enhancement factor of $\enhancemax=10$ (see
  equation [\ref{bigmass}]). For all three distributions, the secondary
  mass is constrained to be less than the total disk mass. } }
\label{fig:shift}
\end{figure} 

\subsection{Continued Accretion}
\label{sec:accrete}

After the initial onset of gravitational instability, any secondary
body that survives can grow larger by accreting additional mass from
the disk. The final mass $M_{\rm f}$ of the object is then given by
\be
M_{\rm f} = \enhance M_{\rm i} \,, 
\ee
where $M_{\rm i}$ is the mass of the secondary at the time of
gravitational instability (from Section \ref{sec:mass}).
{In this paper, we consider two scenarios for continued mass 
accretion, as specified by the distributions of the enhancement
factor $\enhance$.} 

{We first consider} the mass enhancement factor to have a maximum
value $\enhancemax$ = 2. If the growing body accretes more
than its initial mass after the onset of instability, then the
gravitational instability does not strictly speaking determine its
mass. Instead, the object would be formed by a somewhat different
paradigm, where a seed body is formed by instability, but the majority
of the final mass is gathered through accretion {(see below)}. 
In any case, the final mass (in log-space) is related to the initial
mass through the relation
\be
\zeta_{\rm f} = \zeta_{\rm i} + \ln\enhance \,. 
\ee
Since the enhancement factor $\enhance$ is bounded from above, the
added variable $\ln\enhance$ is confined to a relatively small
interval. The resulting distribution for the final mass (given by
$\zeta_{\rm f}$) is thus not overly sensitive to the distribution of
$\ln\enhance$, so we take the distribution to be log-uniform on the
interval [0, $\ln2$]. With this choice, {the mean and width of
the composite mass distribution increase according to} 
\be
\langle \zeta_{\rm f} \rangle = \langle \zeta_{\rm i} \rangle +
{1\over2}\ln2 \qquad {\rm and} \qquad 
\sigma_{\rm f}^2 = \sigma_{\rm i}^2 + {1\over12} (\ln2)^2 \,. 
\ee
{The shift in the mean is small ($\sim0.35$) and the shift in
the variance is even smaller ($\sim0.04$). } 

{Next we consider larger values for the enhancement factor. In
principle, the secondaries could grow until they reach the isolation
mass, where the bodies have accreted all of the mass in an annulus of
width of several $R_H$ centered on the orbit. This mass scale can be
an appreciable fraction of the total disk mass \citep{kratter2016}.
However, continued accretion is constrained by additional
considerations.} With their expected masses, the secondary bodies of
interest most likely exceed the mass threshold for opening a gap in
the background disk \citep{paardekooper}, which limits additional
accretion \citep{xu2024}. {The rate of continued accretion is a
sensitive function of the initial mass \citep{kratter2016}. If the
starting mass is large enough to clear a gap, then continued accretion
is slow. In contrast, small objects can remain embedded and accrete
more rapidly. Finally, for completeness, we note that} the fragments
can lose mass through tidal stripping (e.g., \citealt{nayakshin2010}),
which acts to further limit the value of $\enhancemax$. {These
considerations can be taken into account by making the probability
distribution extend up to $\enhance=\enhancemax=10$, but with
decreasing probability and a cutoff at the disk mass. If we take the
peak of the original distribution to have mass $M_0=15\mjup$, then
this cutoff requires $\zeta<3$ for a disk mass $\mdisk=0.3M_\odot$.
The aforementioned features can be included by using a probability
distribution of the form} 
\be
P(\enhance) = {2 \over \ln\enhancemax} \left[1 -
  {\ln\enhance\over\ln\enhancemax} \right]\,,
\label{bigmass} 
\ee
{which is defined on the interval $[0,\ln\enhancemax]$. For this
distribution for $\enhance$, the mean and width of the composite
distribution become}
\be
\langle \zeta_{\rm f} \rangle = \langle \zeta_{\rm i} \rangle +
{1\over3}\ln10  \qquad {\rm and} \qquad 
\sigma_{\rm f}^2 = \sigma_{\rm i}^2 + {1\over18} (\ln10)^2 \,. 
\ee
{The shift in the mean becomes $\sim0.76$ with a corresponding
shift in the variance $\sim0.29$.} 

{The mass distributions due to continued accretion are illustrated
in Figure \ref{fig:shift}. Note that the minimum secondary mass
remains the same. For the first case, the peak of the distribution
shifts to the right, but retains its shape (except for a small
correction near the cutoff at the mass scale of the disk). For the
second case with $\enhancemax=10$, the rightward shift is larger, and
the cutoff due to the finite disk mass becomes more important.  Note
that we have used a large disk mass ($\mdisk=0.3M_\odot$) and that
smaller disk masses would lead to smaller changes to the SMF.}

{In the limit where most of the mass is gained through accretion,
the initial collapse creates a seed mass $M_0\sim10\mjup$, but
secondary masses are determined primarily by continued accretion
rather than by the gravitational instability. This scenario is
analogous to the core accretion process, where a seed mass of order
$M_0\sim20M_\oplus$ experiences runaway accretion and grows much
larger. The distribution of final masses can be calculated for likely
accretion scenarios with ${\dot M}\propto M^p$ (where $p\sim2/3-4/3$;
see \citealt{adams2021}). In the case of core accretion, the planets
gain most of their mass after core formation, with the final masses
spanning a range of $\sim100$, and the mass distribution approaches a
power-law.} Moreover, the accretion process runs out of time as the
background disk dissipates. In contrast, for gravitational instability
with continued accretion, the secondary mass is limited because the
disk eventually runs out of mass.

\section{Conclusion}
\label{sec:conclude} 

\subsection{Summary of Results}
\label{sec:summary} 

For secondaries forming through gravitational instability in disks, the
mass scale can be defined using the Hill radius, the Bondi radius,
the disk scale height, the fastest growing wavelength for spiral
instabilities, and/or the Jeans length. We have shown that all five
approaches lead to the same mass scale, up to dimensionless factors
of order unity, as expressed in equation (\ref{massgeneral}). The
convergence of these five seemingly different approaches arises
because the circumstellar disk must be unstable, with Toomre $Q$
parameter of order unity, and because the instability process is
intrinsically local (so that environmental effects are described
by local variables such as the mean motion $\Omega$). 

The mass scale of equation (\ref{massgeneral}) corresponds to the
masses that objects would have if they can form through gravitational
instability. In order for secondary formation to occur, the cooling time
must be sufficiently short. We have shown that this cooling constraint
(which has been discussed previously -- see \citealt{rafikov2005},
\citealt{lizano2010}, and others) is equivalent to the concept of
opacity limited fragmentation (also discussed previously --- see 
\citealt{rees1976}). Cooling constraints provide a lower limit to the
masses for objects that can form via gravitational instability, where
the lower limit $M_{\rm min}\sim4-8\mjup$.

The mass available in the circumstellar disk provides an upper limit
to the masses of companions that form through this mechanism. This
constraint implies an upper limit of $M_{\rm max}\sim80\mjup$
{(see Appendix \ref{sec:maxfrag})}. This threshold thus falls
near the boundary between brown dwarfs and true stars. 

With the minimum and maximum masses determined (Section
\ref{sec:minfrag} and Appendix \ref{sec:maxfrag}), the range of
secondary masses is finite, where $M$ varies by a factor of
$\Lambda_T\sim20-30$.  This mass range is much smaller than that
spanned by stars (which have a mass range of a factor of $\sim1500$)
and that of all planets (which have a mass range\footnote{Here we take
the upper mass limit for planets to be $10\mjup$ and the lower limit
to be the mass of Mars.}  corresponding to a factor of at least
$\sim32,000$).  For completeness, note that brown dwarfs have an even
smaller mass range, only about a factor of $\sim8$.

\begin{figure}
\centering
\includegraphics[width=0.80\linewidth]{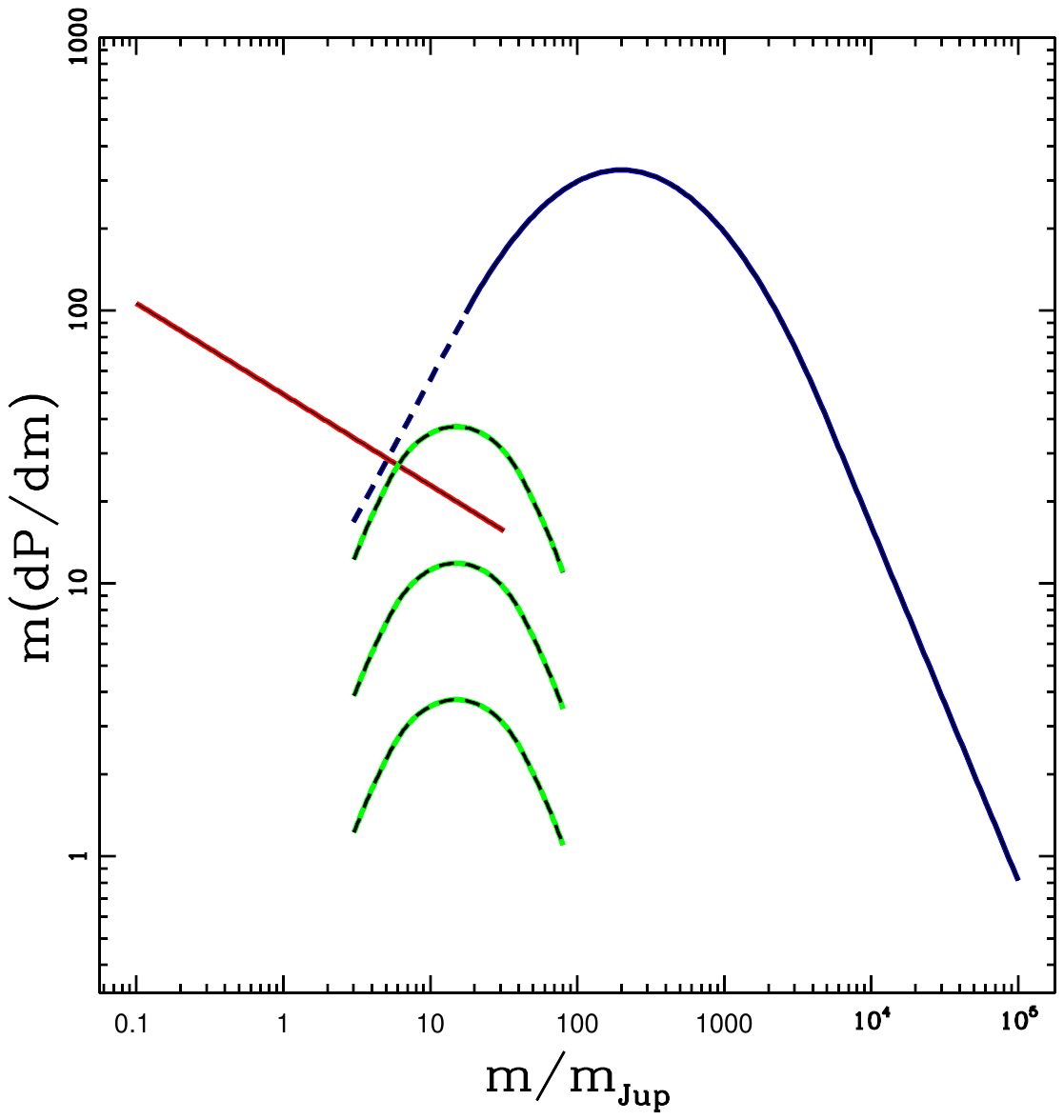}
\vskip-1.30truein
\caption{Comparison of mass functions for different types of
  astronomical objects. The blue curve (right) shows the stellar
  initial mass function, which is taken to have a log-normal form and
  a power-law tail at high masses. The dashed blue curve shows a
  power-law tail for low masses. The red curve (left) represents the
  mass function for giant planets, assumed here to have a power-law
  distribution and an occurrence rate of 25\%.  Three green curves
  show the expected mass function for secondaries formed via
  gravitational instability, from this paper, where the distributions
  are normalized for occurrence rates of 1, 3.16, and 10\% (bottom to
  top). Note that the occurrence rate must be of order 10\% (top green
  curve) or larger in order for gravitational instability to make a
  measurable contribution to the overall population. {Notice also
  that some fraction of the stars are binary, and those secondaries
  have a corresponding distribution of masses (not shown).} }
\label{fig:imf}
\end{figure}  

The resulting mass formula from equation (\ref{massone}) depends on
only three variables $(\eta,v_s,\Omega)$. These three variables, in
turn, depend on the radial location $r$, the stellar mass $M_\ast$,
and the overall luminosity which sets the scale $T_D$ of the disk
temperature distribution. We must combine the dependence of the sound
speed $v_s$ and the rotation curve $\Omega$ on the radius $r$ and the
stellar mass $M_\ast$ so that we are left with independent variables
to sample (see equation [\ref{scalemass}]). These independent
variables can be taken to be $(\eta,T_D,M_\ast,r)$ or their
dimensionless equivalents $(\eta,\beta,\mu,\xi)$.

Since the constituent variables $(\eta,\beta,\mu,\xi)$ are (by
construction) independent, the width of the composite distribution
(the SMF) is the sum of the widths of the individual distributions
(see equation [\ref{totalwidth}]). This result holds for any form of
the distributions of the constituent variables.

The mass of forming secondaries is only weakly dependent on the stellar
mass (again see equation [\ref{scalemass}]). Most of the dependence on
stellar mass scales out of the problem: Larger stellar masses produce
hotter disks (increasing the numerator of the mass formula) but also
produce faster rotation curves (thereby increasing the denominator).
As as result, the two effects nearly compensate, leaving little
dependence on stellar mass. One possibly important issue is that disks
around low mass stars are expected to have smaller total masses (we
expect the mass ratio $\mdisk/M_\ast\lta0.1$). The upper limit for
this channel of secondary formation will thus be smaller for smaller
stars, and that constraint will influence the distribution, {e.g.,
by limiting the amount of continued accretion (see Figure \ref{fig:shift}).} 

In summary, we have constructed a working model for the Secondary Mass
Function (SMF) for objects forming via gravitational instability. If
we take the distributions of the constituent variables
$(\eta,\beta,\xi)$ to be log-uniform, and fix their {\it weighted}
widths to be equal, then the resulting distribution for secondary mass
(see Figure \ref{fig:dist}) has approximately the expected mean,
width, and range. Recent numerical studies \citep{xu2024} find
similar results, where gravitational instabilities lead to the
formation of secondaries with masses in the range 4 -- 100 $\mjup$
with a nearly log-normal distribution. 

To place these results in context, Figure \ref{fig:imf} compares the
stellar initial mass function (blue), the planetary mass function for
core accretion planets (red), and the mass function for gravitational
instability objects from this paper (green). The stellar IMF is taken
to have a log-normal form (with mass scale $m_0=0.20M_\odot=200\mjup$
and width $\sigma$ = 0.69; see, e.g.,
\citealt{scalo1986,adams1996,chabrier2003,sumi2023,defurio2025}),
as well as a power-law tail at high masses (with a slope given by
($m(dP/dm)\sim m^{-1.3}$; \citealt{salpeter1955}). The dashed blue
curve shows extends the stellar IMF with a shallower slope well into
the brown dwarf regime (down to 3 $\mjup$), as indicated by recent
surveys (e.g., \citealt{kirkpatrick2021,defurio2025,kirkpatrick2024}).
The planetary mass function is modeled as a power-law with slope given
by $m(dP/dm)\sim m^{-1/3}$ \citep{cumming2008,fulton2021,meyer2025},
normalized such that the occurrence rate of gas giant planets is 25\%
(for the mass range $m$ = 0.1 -- 10 $\mjup$ and for all orbits). The
three green curves for gravitational instability secondaries (this
work) are normalized with occurrence rates of 10\% (top), 3.16\%
(middle), and 1\% (bottom). Current estimates \citep{vigan2021}
suggest that the occurence rate lies near the bottom on this range.
Note that brown dwarfs --- objects with mass $m\approx10-80\mjup$ ---
can be formed as the high-mass end of the core accretion distribution
(red), the low-mass end of the stellar distribution (blue), and
through the action of disk instability (green). In spite of multiple
formation pathways, however, their total contribution to the
population of astronomical objects is limited {(thereby
contributing to the apparent brown dwarf desert). Notice also that
some fraction of the stellar population has stellar companions, and
that this additional mass distribution is not shown. } 

\subsection{Discussion}
\label{sec:discuss}

This paper has constructed a working theory for the secondary mass
function for secondaries forming through gravitational instability.
However, this treatment represents only a starting point, and a number
of complications should be considered in the future:

{\sl Input Distributions:} The constituent variables that enter into
the mass formula are well-defined physical quantities (e.g., sound
speed, rotation rate), and their typical values and expected ranges
can be estimated to a reasonable approximation.  On the other hand,
the underylying distributions of these variables is not well
known. Fortunately, the three variables $(\eta,\beta,\xi)$ contribute
(mostly) equally to the width of the composite distribution and have
finite ranges that are relatively narrow. These properties tend to
make the composite distribution (the SMF) well behaved, with a single
peak in the center of a modest range of values. As a result, the
simple models presented here are likely to provide a good
approximation for the SMF, but an improved specificiation of the
constituent distributions would be desirable. 

{\sl Magnetic Fields:} The inclusion of magnetic fields
{can affect} the process of gravitational instability.
The additional pressure from magnetic fields inhibits secondary
formation, but the effect of magnetic pressure on the underlying
rotation curve of the circumstellar disk (which becomes subkeplerian)
enhances secondary formation. The two effects essentially cancel 
\citep{lizano2010}. In addition, the forming object must get rid of
its magnetic flux in order to shrink down to planetary size scales.
This requirement leads to a lower limit on the electrical resistivity
of the disk gas. For completeness, we note that one recent study
\citep{deng2021} finds that magnetic fields can lead to the formation
of smaller objects.\footnote{These lower mass condensations survive
because the magnetic fields confine growing density perturbations in
the disk that would otherwise disperse. Note that the \cite{deng2021}
simulations use system parameters that depart from those of
\cite{lizano2010} and others, e.g., the resistivity is different by a
factor of $\sim100$. } In the present context, we except magnetic
fields to hinder the secondary formation process only modestly,
perhaps increasing the lower limit of the mass range, and lowering the
overall occurence rate.

{\sl Migration:} Planets and brown dwarfs can move while they are
forming and after they form. The time scale for gravitational collapse
is of order the orbit time -- both because the cooling time must be
shorter than (or of order) the orbit time and because the time scale
for robust gravitational instabilities is set by the orbit time
\citep{adams1989}. In contrast, the migration time is much longer, and
takes place over many orbits. As a result, if migration leads to an
appreciable change in the secondary mass, the process is one of
accumulation rather than gravitational instability (see above).
Nonetheless, the secondary orbit can change through migration, so the
currently observed orbits might not be the same as the formation
sites.

In order to move an object, the circumstellar disk must have
sufficient mass. As a general rule of thumb, to move a body one
e-folding in semimajor axis, the background disk in the local annulus
(within a factor of ${\rm e}\approx2.718$ of the orbital location)
must have a mass comparable to that of the body. Mass considerations
act to limit migration to only a few e-foldings. Another way to move
objects is through scattering events, although this mechanism requires
additional bodies. In general, the largest objects are scattered
inward, whereas the smaller objects are scattered outward. In
addition, the mass in the `scattering' bodies must be comparable to
the mass of the `scattered' body, so the mass supply once again limits
the opportunities for migration.

{\sl Continued Accretion:} Objects formed through gravitational
instability have a given mass after their initial collapse, but can
gain additional mass through subsequent accretion. This paper
considers this process in a statistical fashion (Section
\ref{sec:accrete}) using the constraint that the starting mass must
provide the majority of the final mass. This constraint limits
the amount of additional accretion to be no more than twice the
initial mass.  However, a conceptually different paradigm could
operate, where gravitational instability provides a seed mass that
continues to grow to much larger values. In this case, most of the
mass is gathered through an accretion process, and this channel of
secondary formation should be studied in greater depth. 

{\sl Outer Disk Boundary:} The radial location where objects can form
is highly constrained: Secondaries cannot form at small semimajor axes
because cooling constraints disallow their survival. This constraint
limits these objects to form at $a\gta100$ AU. On the other hand,
objects cannot form at overly large semimajor axes because the disks
do not extend far enough and the surface density must exceed a minimum
value (equation [\ref{surfcon}]). For example, the recent D-Sharp
observational survey generally shows disk radii $\rdisk\sim100$ AU
\citep{andrews2018}.  Note that it remains possible for some disk
material to reside further out, beyond where disk emission is
detected, but the absence of observations indicates that the amount of
mass at such large distances is (most likely) limited. {Additional
observations are needed to inform this issue. } 

{\sl Wide Companions:} Although circumstellar disks are generally not
observed beyond a few hundred AU, we do see secondary bodies at large
separations, and star forming regions often have non-zero emission at
such distances. Since centrifugally supported circumstellar disks do
not extend to such large radii, however, secondary formation via disk
instabilities does not occur at these locations. The observed
secondary bodies could arise from several different mechanisms: [1]
The objects could be formed at closer distances, and then moved
outward through one of the migration processes described above. [2]
The bodies could be captured by the star from a freely-floating
state. [3] The secondaries in question could be freely-floating bodies
and not actually bound to the nearest star (note that we have not
observed full orbits for these objects). [4] The bodies are `brown
dwarfs' rather than `planets' and form via a `star formation
process'. In this latter case, the collapsing molecular cloud core
(that forms the star and disk) would develop a second condenstation
center, which forms a secondary body at large radius. Note that this
type of behavior must take place (somehow) in order to form wide
stellar binaries (at least those that are not formed via capture). As
a result, the small planet-like bodies in this scenario represent the
tail of the stellar binary distribution (but strictly speaking do not
form via gravitational instability in disks).

{Note that more massive stars tend to have larger disks, both in
terms of mass and radius. A recent survey searching for giant planets
around B stars \citep{delorme2024} discovered two objects with masses
in the range $M=10-15\mjup$ and wide separations $a=290-560$ AU. With
host star masses $M_\ast=6-9M_\odot$, the parental disks could have
extended to these sizes, and the secondary mass scales (which are
largely independent of stellar mass) fall in the expected range for
gravitational instability. The population of such objects, with B star
hosts, provides a promising opportunity to observe the results of the
formation process considered in this paper, including (eventually) the
SMF. }

{\sl Population Contamination:} A related issue is that the mass range
for secondaries formed by gravitational instability overlap with those
formed via core accretion in the approximate range $M=4-10\mjup$.
Moreover, brown dwarfs and stellar binary companions can form readily
(e.g., \citealt{caballero2007}), down to mass scales of order
$\sim20\mjup$. As a result, we strongly suspect that observed
secondaries with masses $M=10-20\mjup$ are good candidates for objects
that form through gravitational instability, but somewhat smaller and
larger bodies might form through alternate mechanisms. It is important
to point out that we do not have a fully consistent and predictive
theory for binary star formation (e.g., see the reviews of
\citealt{duchene2013,offner2023}), and this absence makes it difficult
to separate the various formation mechanisms.

In summary, we find that secondary formation through the action of
gravitational instability in circumstellar disks is highly
constrained. The combined requirements of unstable disks and
sufficient cooling imply that secondaries must form at large radii
$r\sim100$ AU. Since disks do not extend much beyond this radius, the
expected birth location for this mechanism to operate is predicted to
be near $r=100$ AU.  The minimum mass for a forming object is given by
the limit for opacity limited fragmentation, whereas the maximum mass
is determined by the supply of gas in the disk. We thus obtain a
relatively narrow mass range $4\le{M/\mjup}\le80$. The bottom line is
that gravitational instability produces $M\approx10\mjup$ objects at
$r\approx100$ AU. For this population of secondaries, we have
constructed the mass function, which is predicted to be a rather
narrowly peaked (gaussian-like) distribution. Significantly,
reasonable estimates for the distributions of the constituent
variables lead to a SMF that is consistent with current (limited)
observations, as well as extant numerical simulations.  More
specifically, the expected values of the physical input variables
(e.g., the temperatue and rotation curve at 100 AU) lead to the
correct mass scale, and the expected range of values lead to
(approximately) the correct width of the mass distribution. The
framework developed in this paper can be readily generalized to
include more physical effects, many of which are outlined above.
On the observational front, recent and upcoming surveys from the
{\it James Webb Space Telescope} (e.g.,
\citealt{luhman2024,crotts2025})
the {\it Nancy Grace Roman Space Telescope} (e.g.,
\citealt{johnson2020}), and the {\it Gaia} Mission (e.g.,
\citealt{holl2022}) will help determine the populations of secondaries
in the mass range of interest.

\begin{acknowledgments} 

We thank an anonymous referee for useful feedback.  F.C.A. is
supported in part by the Leinweber Institute for Theoretical Physics
at the University of Michigan. A.G.T. acknowledges support from the
Fannie and John Hertz Foundation and the University of Michigan’s
Rackham Merit Fellowship Program.

\end{acknowledgments} 


\appendix

\section{Maximum Fragmentation Mass}
\label{sec:maxfrag}

This Appendix places upper limits on the mass of secondaries that can
form through gravitational instability in disks. The first (but
weakest) constraint is that the forming object must have less mass
than its host disk,
\be
M < M_{\rm disk} \,.
\ee
Since disks have maximum masses of order 1/10 of the stellar mass, and
since we are considering solar type stars, this constraint implies
that the maximum secondary mass is less than about $M_{\rm max}$
$\approx0.1M_\odot=100\mjup$. However, it is highly unlikely for the
forming object to gather up the entire mass of the disk. The upper
limit for gravitational instability objects can thus be taken to be
somewhat smaller, $M_{\rm max}\approx80\mjup$, i.e., near the mass
scale of star/brown dwarf boundary.

The above arugement implies that true stars are probably not formed as
secondary bodies via gravitational instability in disks.  This present
scenario considers the system to consist of a well-defined
circumstellar disk that is centrifugally supported in orbit about an
equally well-defined stellar body. Such systems are subject to the
upper limit for the disk mass $(M_{\rm disk}\lta M_\ast/10$). To evade
this bound, binary formation must take place through the fragmentation
of systems that are less `well-defined', in that they do not separate
cleanly into a star and centrifugally supported disk.

The limits described above are not the least upper bounds on the
secondary mass, i.e., tighter limits can be found {(using a
version of the isolation mass argument)}: Consider the 
scenario where the surface density of the disk has a power-law form
$\Sigma \sim r^{-3/2}$. For this case, the mass enclosed within a
radius $r$ is given by the simple expression 
\be
M_d(r) = M_{\rm disk} \left({r \over \rdisk}\right)^{1/2} \,. 
\ee
An object forming via gravitational instability could in principle
gather up all of the mass within an annulus of the disk. The width
of this annulus is expected to be a multiple $f$ of the Hill radius
$R_H$, so the total mass of the annulus provides an upper limit on
the mass,
\be
M < \int_1^2 2\pi r dr \Sigma(r) \,, 
\ee
where the limits of integration correspond to the width of the
annulus.  Since the Hill radius must be small compared to the radius
itself, otherwise the concept of the Hill radius is invalid, the 
above expression simplifies to the form 
\be
M < M_{\rm disk} f {R_H \over \rdisk} =
M_{\rm disk} f {r \over \rdisk} \left({M\over3M_\ast}\right)^{1/3}\,. 
\ee
Solving for the mass $M$ and using the inequality $r<\rdisk$,
we find the limit
\be
M < {f\over\sqrt{3}} { M_{\rm disk}^{3/2}\over M_\ast^{1/2}} \,. 
\ee
If we evaluate this constraint using $f=3.5$ (Kratter et al. 2010),
$M_{\rm disk}/M_\ast$ = 0.1, and $M_\ast=1M_\odot$, the right hand
side of the above equation becomes $\sim64\mjup$. Allowing for
somewhat more massive disks, the mass scale can approach the brown
dwarf limit, so we again find the approximate upper bound 
\be
M < M_{\rm max} \approx 80 \mjup \, . 
\ee

\section{Alternate Probability Distributions} 
\label{sec:altprob}

In Section \ref{sec:distrib}, the probability distributions of the
constituent variables are taken to be log-uniform (equation
[\ref{unilog}]) and the resulting composite distribution (the SMF) has
a nearly log-normal form (Figure \ref{fig:dist}). Although this result
is expected in the limit of large number $N$ of independent variables
\citep{richtmyer1978}, we find that the convergence is rapid, with
gaussian-like forms arising for $N$ = 3 or 4. This Appendix explores
how this result depends on the assumption that the probability
distributions have a particular form. We thus consider alternate
constituent distributions, and show that the resulting composite
distribution is relatively close to gaussian in shape. 

For the sake of definiteness, this treatment works in terms of
logarithmic variables (see Section \ref{sec:distrib}). In this case,
the composite variable $\zeta$ = $\ln(M/M_0)$ that specifies the mass
of the secondary is the sum of $N$ constituent variables,
$$\zeta = \sum_{k=1}^N x_k \,. $$
The composite distribution has a finite width, with variance
$\sigma_\zeta^2$ = 1, so that the individual distributions must have
$\sigma_k^2=1/N$. Note that without loss of generality, we can shift
the variables so that they have zero mean $(\langle{x}\rangle=0)$.

\begin{figure}
\centering
\includegraphics[width=0.80\linewidth]{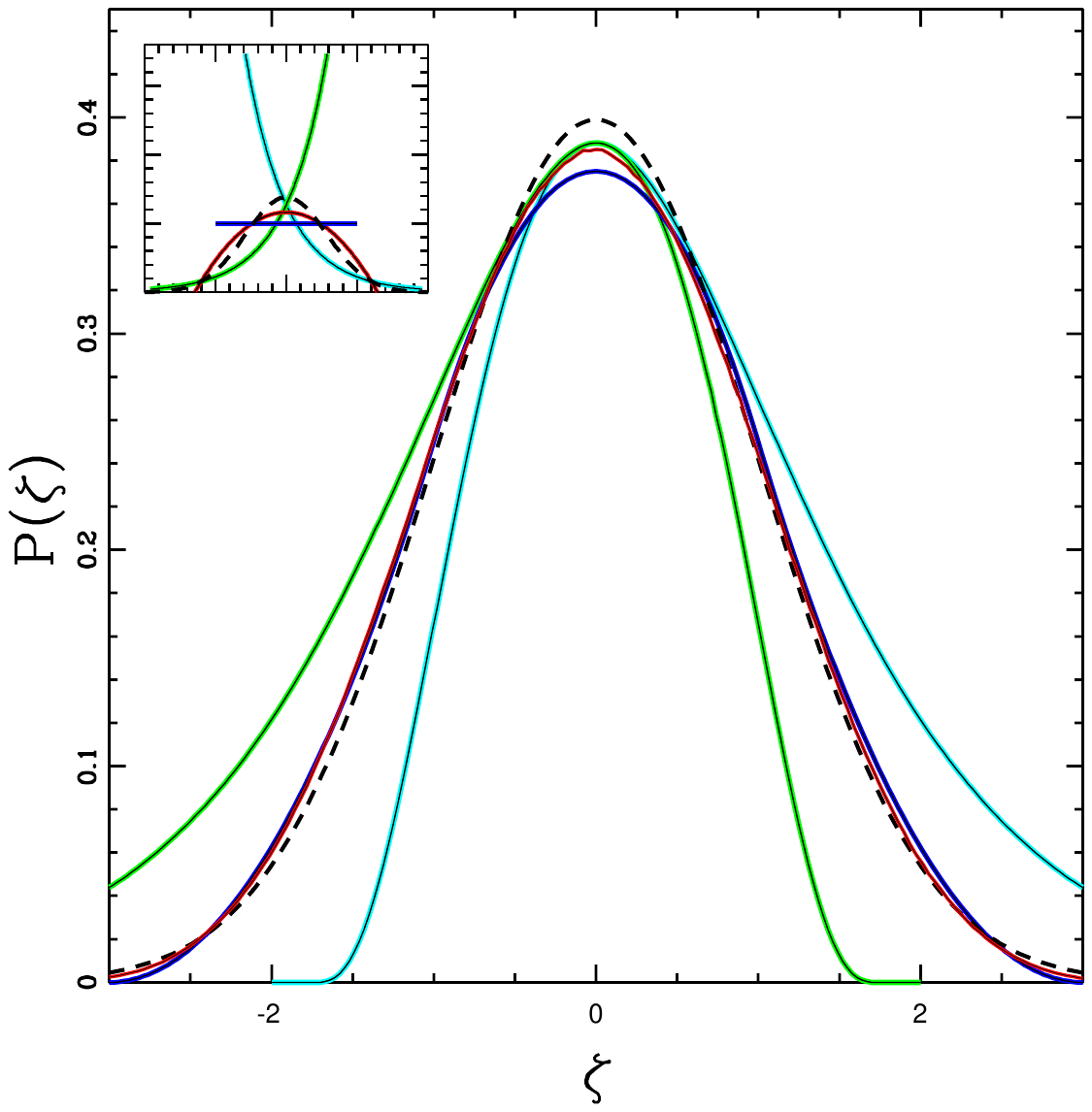}
\vskip-1.30truein
\caption{Comparison of composite distributions using different forms
  for the distributions of the constituent variables. The solid curves
  correspond to the sum of three random variables chosen from
  disributions that are log-uniform (blue), inverse quadratic (red),
  decaying exponential (cyan), and growing exponential (green). The
  black dashed curve shows a gaussian distribution for comparison.
  The inset shows the constituent distributions. The composite
  distributions have the same variance, but those resulting from
  exponentially varying constituent distributions are tilted with
  respect to the gaussian.}
\label{fig:altprob}
\end{figure}  

As a contrast with the benchmark choice of a uniform distribution, we
consider an exponential distribution for the constituent variables. If
the variable $x$ has a minimum value, but no maximum, then the
distribution can be written in the form 
$$p(x) = \gamma \exp[-\gamma x - 1]\,, $$
where the range (support) is given by $x\in[-1/\gamma,\infty)$. 
The distribution has zero mean and variance $\sigma_1^2=1/\gamma^2$,
so we must take $\gamma=\sqrt{N}$. If we consider the sum of $N$
random variables with the distribution given above, the composite
variable has a gamma distribution, which can be written
$$f_N (\zeta) = {N^{N/2} (\zeta + \sqrt{N})^N \over (N-1)!}
\exp[-\sqrt{N}\zeta - N]\,, $$
with the range given by $\zeta\in[-\sqrt{N},\infty)$. Note that this
range is infinite, whereas the constituent variables of interest have
a (relatively narrow) finite range. In practice, one should impose a
cutoff. For purposes of this exploration, however, we want to
determine how far the composite distribution can depart from a
gaussian form, so the infinite tail of the distribution is retained,
thereby providing a worst-case scenario. One can also consider a
growing exponential, which leads to the mirror image of the above
expression with the range $\zeta\in(-\infty,\sqrt{N}]$.

As an alternate possibility, one can consider a more compact
distribution for the constituent variables.\footnote{Although a
gaussian distribution is an obvious choice, it is well known that the
convolution of gaussian distributions leads to another gaussian.
Since we are interested in cases that could depart from a gaussian
composite, we choose an alternate form.} Here we choose an
inverse quadratic distribution of the form 
\be
p(x) = {3\over4b^3} (b^2 - x^2)\,.
\label{pquad} 
\ee
As written, the distribution is normalized and has zero mean. The
parameter $b$ can be chosen to obtain the proper variance of the
composite distribution, so that the present application requires
$b^2=5/N$. The composite distribution for a sum of variables
distributed according to equation (\ref{pquad}) can be carried out
analytically, but results in rather convoluted expressions (not
shown).

Figure \ref{fig:altprob} shows the resulting composite distributions
for the sum of three random variables chosen from the distributions
presented above. Since we are interested in finding the largest
possible departures from a gaussian form for the composite
distribution, we have taken the smallest value of $N=3$. The four
colored curves correspond to the different choices of input
distributions, including uniform (blue), inverse quadratic (red),
decaying exponential (cyan), and growing exponential (green). The
dashed black curves shows a gaussian distribution for comparison. The
corresponding input distributions (with the same color coding) are
shown in the inset figure in the upper left. The composite
distributions resulting from uniform and inverse quadratic input
distributions are difficult to distinguish from a gaussian form.  The
composite for the decaying exponential (cyan curve) has the same
width, but is skewed with a tail on the right and a deficit on the
left. The composite for the growing exponential displays the opposite
behavior, with a tail on the right (green curve). Note that these
exponential input distributions are more extreme than those expected
in nature: The exponentials have an infinite range, rather than a
limited range of values, and these results are showns for the fewest
possible number of variables. Even in this extreme case, however, the
composite shows only a modest departure from a gaussian form,
specifically the skewness described above.

\end{document}